\documentclass[12pt,a4paper]{article}
\usepackage[dvips]{color,graphicx}
\usepackage{amsfonts}
\usepackage{amssymb}
\usepackage{latexsym}
\usepackage[latin1]{inputenc}

\newcommand{\lw}[1]{\smash{\lower2.ex\hbox{#1}}}

\begin{document}

\title{Homothetic motions and Newtonian cosmology}

\author{Xavier Ja\'en\thanks{Dept. de F\'isica i Enginyeria Nuclear, Universitat Polit\`ecnica de Catalunya, Spain, e-mail address: xavier.jaen@upc.edu} and Alfred Molina\thanks{Dept. F\'{\i}sica Fonamental, Institut de Ci\`encies del Cosmos, Universitat de Barcelona, Spain, e-mail address: alfred.molina@ub.edu}}

\maketitle

\begin{abstract}
We consider the homothetic motion group. We construct a homothetic covariant Newtonian gravitation theory which unifies inertial homothetic forces and gravitational fields. This is achieved through an equivalence principle based on a local homothetic frame of motion. As a consequence, we can obtain a coherent Newtonian cosmology which admits a cosmological principle and leads to the  Friedman equations for a dust universe. Finally we prove that this gravity theory can be obtained as the non-relativistic limit of a class of metrics in General Relativity. The Friedmann-Lema\^\i tre-Robertson-Walker (FLRW)
metric and its limit are also studied.   
\end{abstract}

\section{Introduction}
As is well known, Newtonian laws of mechanics are also valid in non-inertial frames, related between themselves by a transformation of the group of rigid motions, provided that inertial forces are included. 

In a previous paper \cite{Jaen-Molina}, we set up a theory of generalized Newtonian gravitation based on a local velocity field as the vector potential. That theory of gravity treats both the inertial force fields and the Newtonian gravitational field in a unified way. 

An interesting result of the theory is that the same vector field, $\vec v_g$, occurs in the description of spherically symmetric gravitational solutions, both in generalized Newtonian gravitation and in General Relativity for the Schwarzschild metric written in Painlev\'e-Gullstrand\cite{Painleve} coordinates.

The aim of the present paper is, by following an analogous procedure, to obtain a Newtonian cosmology equivalent to relativistic FLRW cosmology and which could be seen as a non-relativistic limit of it. 

Usually in Newtonian mechanics the reference frames are rigid bodies: rigid rods are used to measure space  and a (universal time) clock to measure time. If for a reference frame, however, the axes are rescaled isotropically, then all the measuring rods and objects in that frame are rescaled by the same factor. An observer in such a frame is not aware of the rescaling and then it is pushed 
to use the Euclidean metric.  The convenience  of using a constant curvature 
space metric instead of a (possible) time dependent one has been pointed out by L.Bel\cite{LB} in the context of General Relativity. Can this new kind of frame be used also as a possible physical observer to be added to the usual family of non-inertial observers? What are the inertial forces that then appear?

We are now going to extend the rigid group of motions to a larger one that includes isotropic scaling as a new transformation of the group; this is known as homothetic motions.  The fact that the class of homothetic motions has a group structure allows us to use it as a covariant group of the theory. If we are able to find a homothetic-covariant formulation of Newtonian gravitation we will be able to answer the two preceding questions.

The paper is organized as follows: in section \ref{sec2} we define homothetic motion frames of reference. For these frames we find the equation of motion for particles and the continuity equation for matter. We define the inertial fields for homothetic motions and provide the field equations. In section \ref{sec3} we characterize gravity through velocity field and scalar field potentials, 
according to the homothetic inertial fields found in section \ref{sec2}. Then we  define the local homothetic reference frame and state the corresponding equivalence principle, which we use to find the equation of motion of particles and the continuity equation. 
In section \ref{sec4}, again through use of the equivalence principle, we give the Lagrangian formulation for particles subject to a gravitational field. In section \ref{sec5} we provide non-relativistic equations for the gravitational potentials that, together with the equation of motion for particles and the continuity equation, form the base of the homothetic covariant Newtonian gravitation theory.
Section \ref{secNC} is devoted to applying the Cosmological Principle to this Newtonian theory which as a result leads to the Friedman equations for a dust universe. In section \ref{sec7} we compare generalized Newtonian gravitation with a non-relativistic limit of General Relativity.
Finally in section \ref{sec8} we analyze FLRW spacetime.

\section{Equations of motion and continuity equation. Inertial force fields\label{sec2}}
\subsection{Homothetic reference frames}
Let us assume that we have two reference frames: 
 $\mathcal{K}$ and $\mathcal{K}^\prime$.
Let  $O$ be the origin of $\mathcal{K}$, whose axes are aligned according to the orthonormal base $\{\vec\varepsilon_i \}_{ i=1\ldots 3}$, and let $O'$ be the origin of $\mathcal{K}^\prime$, whose axes are aligned with the orthogonal base $\{\vec e_i(t) \}_{ i=1\ldots 3}$. Assume that $\overrightarrow{OO'}= X^i(t) \vec\varepsilon_i$. 

As both bases are orthogonal, $\vec\varepsilon_i\cdot\vec\varepsilon_j =  \delta_{ij}\,, \quad \forall t $, $\vec e_i(t)\cdot\vec e_j(t) = H(t)^2 \delta_{ij}$\, from which it follows that:
\begin{equation}\vec e_j(t) = H(t) R^i_j(t)\,\vec\varepsilon_i 
\label{eq0}
\end{equation}
where $R^i_j(t)$ is an orthogonal matrix. 

A point $P$ can be referred to both frames, as $\vec x$ in $\mathcal{K}$ and as $\vec y$ in $\mathcal{K}^\prime$, and we have that $\vec x = \vec X + \vec y $ which, writing  the coordinates explicitly, reads: 
\begin{equation} \label{eq1a}
x^i\vec\varepsilon_i =X^i\vec\varepsilon_i +y^j\vec e_j= ( X^i(t)+H(t) R^i_k(t)\, y^k )\vec\varepsilon_i 
\end{equation}
\begin{equation} 
x^i = X^i(t)+H(t) R^i_k(t)\, y^k \label{eq1}
\end{equation}
It is straightforward to see that this set 
of transformations from $x^i$ to $y^j$ form a group.
We say that two reference frames are homothetic if the coordinates of their points are related through transformation such as (\ref{eq1}).
\subsection{Equation of motion}
Now we give a detailed derivation of the transformation of the velocity and acceleration of a mass point $P$ as seen from two homothetic reference frames.
By differentiation of (\ref{eq0}) with respect to time we have: 
\begin{equation}	
\frac{d\vec e_i }{dt} = (H\dot{R}_i^m  + \dot{H}R_i^m )\vec \varepsilon _m  = \vec\Omega\times\vec e_i  + \frac{\dot H}{H}\vec e_i 
\label{eq2}
\end{equation}			
where
$$
\vec\Omega(t) = \frac{1}{2}\sum\limits_j^{} {R_j^k \dot R_j^l  }\vec\varepsilon _k\times\vec\varepsilon_l
$$	
and $\times$ is the usual vector product in Euclidean space\footnote{From now on, in all the space metric computations we are going to use the Euclidean metric $\vec \varepsilon _i \cdot\vec \varepsilon _j \; = \delta _{ij}.$}.

By differentiation of  (\ref{eq1a}) with respect to time and using (\ref{eq2}) we obtain:
\begin{equation}
\vec v = V^i\vec\varepsilon_i+\dot y^i \vec e_i  + y^i \frac{d\vec e_i }{dt} =\vec V+\vec w +\vec\Omega\times\vec y +\frac{\dot H}{H}\vec y\label{eq3}
\end{equation}
where $V^i = \dot X^i$ and $\vec w =\dot y^i\vec e_i(t)$.
	
By differentiation with respect to $t$ again, we obtain the acceleration of $P$, $\vec a$, in the reference frame
$\mathcal{K}$ when in the reference frame $ \mathcal{K'}$ its velocity is $\vec w$ and its acceleration is $\vec b = \ddot y^i \vec e_i$
\begin{eqnarray}
 & &\hspace*{-3em}\vec a =  \frac{dV^i}{dt}\vec \varepsilon _i  + \ddot y^i \vec e_i  + \dot y^i \frac{d\vec e_i }{dt} +\dot y^i\vec \Omega  \times \vec e_i  + \frac{\dot H}{H}\dot y^i \vec e_i  + y^i\dot{\vec\Omega}\times\vec e_i  + y^i \vec\Omega\times \frac{d\vec e_i}{dt}\nonumber \\
& &  \hspace*{4em}+\frac{d}{dt}\left( \frac{\dot H}{H} \right)y^i \vec e_i = \nonumber   \\ 
  & & \hspace*{-1em} \vec A + \vec b + 2\vec \Omega  \times \vec w + 2\frac{\dot H}{H}\vec w + \dot{\vec\Omega} \times \vec y +2\frac{\dot H}{H}\vec \Omega  \times \vec y + \vec \Omega\times(\vec\Omega\times\vec y) + \frac{\ddot H}{H}\vec y \label{eq4}
\end{eqnarray}		
where $\vec A = \frac{{dV^i }}{{dt}}\vec\varepsilon _i  = \frac{{d\vec V}}{{dt}}$ is acceleration of the origin $O'$ of reference frame $ \mathcal{K'}$  with respect to the frame  $\mathcal{K}$.

The field of velocities of the rest points, $\dot y^i  = 0$, in  $ \mathcal{K'}$ as seen from the reference frame $ \mathcal{K}$ are $\vec v_0 (\vec x,t)$ and can easily be obtained from equations (\ref{eq1}) and (\ref{eq3}):
\begin{equation}
\vec v_0 (\vec x,t) = \vec V(t) + \vec \Omega (t) \times (\vec x - \vec X(t)) + \frac{{\dot H}}
{H}(\vec x - \vec X(t))\label{eq5}
\end{equation}
Then from (\ref{eq5}):
$$
 \frac{1}
{2}\vec \nabla  \times \vec v_0=\vec \Omega (t),\quad  \vec \nabla \cdot\vec v_0=3 \frac{\dot H}
{H}  
$$	
If  we have a free particle in the inertial system, $\vec a=0$, then from (\ref{eq4}) the acceleration in the non-inertial frame $\vec b$ is: 
$$
\vec b =  - \vec A - 2\vec \Omega  \times \vec w - 2\frac{{\dot H}}
{H}\vec w - \dot{\vec\Omega}  \times \vec y - 2\frac{{\dot H}}
{H}\vec \Omega  \times \vec y - \vec \Omega  \times (\vec \Omega  \times \vec y) - \frac{{\ddot H}}
{H}\vec y
$$
This equation gives us the inertial fields, i.e., in the non-inertial frame we can write the equation of motion 
by adding the inertial fields
$$
\vec b = \vec g_I  + \vec w \times \vec \beta _I  + \eta _I \vec w
$$
where $\vec g_I ,\vec \beta _I ,\eta _I $ are the inertial fields which can be derived from two potentials, $\vec v_0(\vec x,t)$, defined in (\ref{eq5}) and $H(t)$, as: 
\begin{eqnarray}
& & \vec g_I  := \vec \nabla \left(\frac{\vec v_0^2}{2} \right) - \frac{\partial\vec v_0 }{\partial t} - \frac{2\dot H}{H}\vec v_0,\nonumber \\ && \vec \beta _I := 2\vec \Omega  = \vec \nabla  \times \vec v_0,\quad\eta _I  :=  - \frac{{2\dot H}}{H}\label{eq6}
\end{eqnarray}
From the definition of the inertial fields $\vec g_I$ and $\vec \beta _I$ in terms of the potentials $\vec v_0(\vec x,t)$ and $H(t)$, it is easy to prove that they fulfill the following field equations:
\begin{eqnarray}
&&
\vec \nabla  \times \vec g_I+\frac{\partial\vec \beta_I}{\partial t}-\eta_I\vec\beta_I=0,\quad \vec \nabla  \cdot\vec \beta_I=0  \label{eq7a}
\end{eqnarray}
Then from (\ref{eq5}), (\ref{eq6}) plus the fact that $H$ is only a function of time, we have that:  
\begin{eqnarray}
&&  \vec \nabla  \cdot \vec g_I=\frac12\vec\beta_I^2+\frac32(\dot\eta_I-\frac12\eta_I^2),\quad \vec \nabla  \times\vec \beta_I=0, \quad \vec \nabla H=0 .\label{eq7b}
\end{eqnarray}

When we substitute (\ref{eq6}) into (\ref{eq7b}) and write $\vec v_I$ and $H_I$ instead of $\vec v_0$ and $H$, we obtain a set of seven equations for the inertial potential fields $\vec v_I(\vec x,t)$ and $H_I(\vec x,t)$ whose  solution is $\vec v_I(\vec x,t)=\vec v_0(\vec x,t)$ , as defined in (\ref{eq5}), and $H_I(\vec x,t)=H(t)$.  This is a generalization  of the corresponding equations we obtained in the previous paper \cite{Jaen-Molina} to include the homothetic field $H_I$.
Now, including these inertial fields, 
the laws of mechanics are invariant for the group of homothetic motions, this is a first step to achieve the non-relativistic homothetic covariance principle, slightly more general than rigid covariance stated in \cite{Jaen-Molina}.
\subsection{Continuity equation}
Let us assume that we have a fluid with a density $\rho(\vec x,t)$ in the inertial frame $\mathcal{K}$, and with velocity field $\vec v(\vec x,t)$. The mass conservation equation reads:
\begin{equation}
 \frac{\partial \rho(\vec x,t)}{\partial t} + \vec\nabla_{\vec x}\cdot(\rho(\vec x,t)\vec v)=0. \label{eq15}
\end{equation}
We can  write the continuity equation in the non-inertial frame $\mathcal{K}^\prime$ . First we have: 
$$\frac{\partial \rho(\vec y,t)}{\partial t} =(\vec v_0\cdot \vec\nabla_{\vec x})\rho(\vec x,t)+\frac{\partial \rho(\vec x,t)}{\partial t}\quad\mbox{and}\quad \vec v=\vec v_0+\vec w
$$
where $\vec w(\vec y,t)$ is the velocity field in $\mathcal{K}^\prime$ and $\vec v_0$ is the non-inertial frame velocity with respect to the inertial frame. Now, the left-hand side of equation (\ref{eq15}) becomes:

$$\frac{\partial \rho(\vec y,t)}{\partial t} +\frac{\partial(\rho w^i)}{\partial x^i}+\rho\frac{\partial v_0^i}{\partial x^i}=\frac{\partial \rho(\vec y,t)}{\partial t} +\frac{\partial(\rho w^j)}{\partial y^j}+3\rho\frac{\dot H}{H}$$
and we can now write the mass conservation equation in the frame $\mathcal{K}^\prime$ as:
\begin{equation}
\frac{\partial \rho(\vec y,t)}{\partial t} +\vec \nabla\cdot(\rho \vec w)+3\rho\frac{\dot H}{H}=0\label{eq16}
\end{equation}
Let us point out that again the product in the frame $\mathcal{K}^\prime$ is performed using the Euclidean metric $\delta_{ij}$.
The equation (\ref{eq16}) can also be written as: 
\begin{equation}
 \frac{\partial \rho'(\vec y,t)}{\partial t} + \vec\nabla_{\vec y}\cdot(\rho'(\vec y,t)\vec w)=0, \label{eq15a}
\end{equation}
where $\rho'=H(t)^3\rho$.
\section{Equivalence principle. Equations of motion and continuity in a gravitational field\label{sec3}}
\subsection{Local frame}
In a similar way as we did in \cite{Jaen-Molina}, we  define gravitational potentials  $\vec v_g(\vec x,t)$ and $H_g(t)$\footnote{For simplicity we reject from the very beginning  a possible space dependence of $H_g$ at a  non-relativistic level, this election will be justified in section \ref{sec7}} as a way to unify the description of gravity and inertial fields. In \cite{Jaen-Molina} this step was achieved by considering $\vec v_g$ as a potential for gravity. Now we add to this vector potential a new scalar potential, $H_g$, in order also to include homothetic inertial fields in the formulation.

At  each point $\vec x$ and for each $t$ we can define a local homothetic frame $\mathcal{K}^\prime$ whose axes are aligned with the orthogonal base $\{\vec e_i(t) \}_{ i=1\ldots 3}$ in such a way that $\vec e_i(t)\cdot\vec e_j(t) = H_g(t)^2 \delta_{ij}$, which are rotating with angular velocity $\vec\Omega_g(\vec x, t)=1/2\vec \nabla\times \vec v_g$,  and its origin, $\vec y=0$, is moving with a velocity  $\vec v_g(\vec x,t)$.

From the definition of $\mathcal{K}^\prime$ and (\ref{eq5}) we have that the velocity and expansion fields, $\vec v_{(\mathcal{K}^\prime)}$ and $H_{(\mathcal{K}^\prime)}$,  that describe the motion of $\mathcal{K}^\prime$ as seen from $\mathcal{K}$ can be expressed as: 

\begin{equation}
\vec v_{(\mathcal{K}^\prime)}(\vec x+\vec y,t) = \vec v_g(\vec x,t) + \vec\Omega_g(\vec x, t) \times \vec y + \frac{{\dot H_g(t)}}{H_g(t)}\vec y ,\quad H_{(\mathcal{K}^\prime)}=H_g(t)
\end{equation}
where $\vec x$ is the position of the origin of $\mathcal{K}^\prime$ and  $\vec y$ is the position of points at rest with respect to $\mathcal{K}^\prime$. 
It is interesting to note that $\vec \nabla_{\vec x}\times \vec v_g=\vec \nabla_{\vec y}\times \vec v_{(\mathcal{K}^\prime)}$ but $\vec \nabla_{\vec x}\cdot\vec v_g$ is usually different from $\vec \nabla_{\vec y}\cdot\vec v_{(\mathcal{K}^\prime)}=3\frac{{\dot H_g(t)}}{H_g(t)}$.
\subsection{Equation of motion}
We thus state our {\em non-relativistic homothetic equivalence principle} for the motion of a particle:\\
\textit{A particle at $(\vec x,t)$ subject to the gravitational interaction, described by $\vec v_g (\vec 
x,t)$ and $H_g(t)$, has no acceleration relative to the local frame $\mathcal{K}^\prime_{(\vec x,t)}$.}

This equivalence principle means that the particle feels a gravitational field in $\mathcal{K}$  in such a way that, with respect to the  local system $\mathcal{K}^\prime$, it has no acceleration; i.e., $\vec b=0$. The acceleration in $\mathcal{K}$ where the gravitational field that acts is from (\ref{eq4}) and $\vec y=0$, we have:
$$ 
\vec a = \vec A + 2\vec \Omega_g  \times \vec w + 2\frac{\dot H_g}{H_g}\vec w
$$	
where the acceleration of the  origin $\vec A$ is: 
$$\vec A= \frac{\partial \vec v_g }{\partial t}+(\vec v_g\cdot\vec \nabla) \vec v_g=
 \vec \nabla \left( \frac{\vec v_g^2 }{2} \right) + \frac{\partial \vec v_g}{\partial t} + 2\vec \Omega_g \times \vec v_g
$$	
and using that $\vec v = \vec v_g  + \vec w$ we obtain: 
$$
\vec a = \vec \nabla \left( \frac{\vec v_g^2 }{2} \right) + \frac{\partial \vec v_g }{\partial t} - 2\frac{\dot H_g }{H_g}\vec v_g  + 2\Omega _g  \times \vec v + 2\frac{\dot H_g}{H_g}\vec v
$$
and, as we did before with  the inertial fields associated with the potentials $\vec v_g$ and $H_g$, we can define the gravitational fields:
\begin{eqnarray}
&&\vec g := \vec \nabla \left( \frac{\vec v_g^2 }
{2} \right) + \frac{\partial \vec v_g }{\partial t} - 2\frac{\dot H_g}{H_g}\vec v_g, \nonumber\\ &&\vec \beta  :=  - 2\vec \Omega _g  =  - \vec \nabla  \times \vec v_g, \quad \eta  :=  2\frac{\dot H_g }{H_g}\label{eq9}
\end{eqnarray}
The acceleration field has changed with respect to that we  found in \cite{Jaen-Molina} by the term containing the vector $\vec v_g$ and the scalar $H_g$ potentials. The rotation field remains the same and a new field, $\eta$, due to the isotropic expansion arises. Then we can write
the acceleration in the gravitational field as:
\begin{equation}
\vec a = \vec g + \vec v \times \vec \beta  + \eta \vec v\label{eq10}
\end{equation}	
where the first two terms formally coincide with the expression given in \cite{Jaen-Molina} and the last term is due to the expansion.
\subsection{Continuity equation}
In an analogous way, we can find the continuity equation  in a gravitational field $\vec v_g (\vec 
x,t)$ and $H_g(t)$. In this case, the equivalence principle can be stated thus:\\
\textit{A moving mass of fluid described by a density $\rho$ and a velocity field $\vec v$ with respect to $\mathcal{K}$ in a neighborhood around $(\vec x,t)$ 
and subject to the gravitational interaction, described by $\vec v_g (\vec x,t)$ and $H_g(t)$, is conserved relative to the local system $\mathcal{K}^\prime_{(\vec x,t)}$.} 

This means that in the local frame $\mathcal{K}^\prime_{(\vec x,t)}$ the continuity equation is:
\begin{equation}
 \frac{\partial \rho(\vec y,t)}{\partial t} + \vec\nabla_{\vec y}\cdot(\rho(\vec y,t)\vec w)=0,\label{eq10a}  
\end{equation}
where $\vec w $ is the velocity field of the fluid relative to $\mathcal{K}^\prime_{(\vec x,t)}$. Now by performing the transformation from $\mathcal{K}^\prime_{(\vec x,t)}$ to $\mathcal{K}$, so taking into account in (\ref{eq10a}) that $\vec w=\vec v-\vec v_{\mathcal{K}^\prime_{(\vec x,t)}}$, we will arrive at the expression of  the continuity equation in the frame $\mathcal{K}$  now subject to the gravitational field $\vec v_g (\vec x,t)$ and $H_g(t)$:   

\begin{equation}
\frac{\partial \rho(\vec x,t)}{\partial t} +\vec \nabla\cdot(\rho(\vec x,t) \vec v(\vec x,t))-3\rho \frac{\dot H_g }{H_g}=0\label{eq16g}
\end{equation}
\section{Lagrangian for the non-relativistic generalized gravitational interaction\label{sec4}}
We are going to find a Lagrangian formulation for the motion of a test particle subject to a gravitational field. We know that in the inertial reference frame $\mathcal{K}$, gravity is described by the local velocity field $\vec v_g (\vec x,t)$ and the scalar field $H_g(t)$. As our non-relativistic equivalence principle states,  in the local non-inertial reference system $\mathcal{K}^\prime_{(\vec x,t)}$ the particle has no acceleration. A good candidate for a Lagrangian in $\mathcal{K}^\prime_{(\vec x,t)}$ following the equivalence principle should be the free  particle Lagrangian, i.e., $ L=m \vec{w}^2/2$, where $\vec w$ is the velocity of the particle relatively to $\mathcal{K}^\prime_{(\vec x,t)}$. 
Now, if we write this Lagrangian in terms of the velocity in the inertial system $\mathcal{K}$, we must use that for $\vec y=0$ we have $\vec w=\vec v-\vec v_{\mathcal{K}^\prime_{(\vec x,t)}}=\vec v-\vec v_g$  so in the inertial frame we obtain:
\begin{equation}
L = \frac{1}{2}m\frac{1}{H_g^2}(\dot{\vec x} -\vec v_g)^2 \label{eq16b} 
\end{equation}
Using this Lagrangian we can find the equations of motion: 
$$
\frac{\partial L}{\partial \dot{\vec x}} = m\frac{1}{H_g^2}\left( {\dot{\vec x} - \vec v_g} \right),
$$
$$
\frac{d}{dt}\frac{\partial L}{\partial\dot{\vec x}} = - m\frac{2\dot H_g}{H_g^3}\left( \dot{\vec x} - \vec v_g \right) + 
m\frac{1}{H_g^2}\left(\ddot{\vec x} - \frac{d\vec v_g}{dt}\right),
$$
$$
\frac{\partial L}{\partial \vec x}= \frac{1}
{2}m\frac{1}{H_g^2}\vec \nabla \left( {\vec v_g^2  - 2\,\dot{\vec x}\cdot\vec v_g} \right).
$$
Using:
$$
\frac{d\vec v_g}{dt} = (\dot{\vec x}\cdot \vec \nabla )\vec v_g + \frac{\partial \vec v_g}{\partial t}
$$  
we can write the Lagrange equations as:
\begin{equation}
\ddot{\vec x} = \vec\nabla\left(\frac{\vec v_g^2}{2}\right) + \frac{\partial\vec v_g}{\partial t} - 2\frac{\dot H_g}{H_g}\vec v_g -\dot{\vec x}\times (\vec \nabla\times \vec v_g ) + 2\frac{\dot H_g}
{H_g}\dot{\vec x}\label{eq9a}
\end{equation}
which is the same result we obtain in (\ref{eq9}) and (\ref{eq10}).
\section{Non-relativistic field equations\label{sec5}}
In \cite{Jaen-Molina} we introduced the non-relativistic field equation which are a modification of the usual Newtonian field equations by adding  rigid motion invariance, and we showed that the limit of the relativistic field equation coincides with them.  We now proceed in a analogous way. 

The fields $\left\{ {\vec g,\vec \beta ,\;\eta } \right\}$, (\ref{eq10}), can be related with the potentials $\vec v_g$ and $H_g$ through:
\begin{equation}
\vec g := \vec \nabla \left(\frac{\vec v_g^2}{2} \right) + \frac{\partial \vec v_g}{\partial t} - 2\frac{\dot H_g}{H_g}\vec v_g,\quad 
\vec \beta  :=  - 2 \vec \Omega _g  =  - \vec \nabla  \times \;\vec v_g, \quad \eta  := 2\frac{\dot H_g}{H_g}\label{eq10b}
\end{equation}					
from the field definitions we have:
\begin{equation}
\vec \nabla\times \vec g =  - \frac{\partial \vec \beta }{\partial t} + \eta \vec \beta ,\quad \vec \nabla\cdot\vec \beta  = 0
\label{eq11}
\end{equation}						
and for the source equations we take into account (\ref{eq7b}) and the fact that for $H_g=1$, the equations must coincide with those presented in \cite{Jaen-Molina}. We take:
\begin{equation}
\vec \nabla \cdot\vec g - \frac{1}{2}\vec \beta ^2  - \frac{3}{2}\left( {\dot \eta  - \frac{{\eta ^2 }}
{2}} \right) =  - 4\pi G\rho, \quad\vec \nabla  \times \vec \beta=0,\quad \vec \nabla H_g=0
\label{eq12}
\end{equation}

When we substitute (\ref{eq10b}) into (\ref{eq12}), we obtain a set of seven equations for the potentials $\vec v_g (\vec x,t)$ and $H_g (\vec x,t)$.

Homothetic motion is the solutions of the source-free  field equations, i.e., from (\ref{eq10b},\ref{eq12}) with  $\rho=0$, we have the solution: 
\begin{eqnarray}
\vec v_g (\vec x,t)&=&\dot{\vec X}(t) + \vec \Omega (t) \times (\vec x - \vec X(t)) + \frac{{\dot H}}{H}(\vec x - \vec X(t)) \nonumber\\
H_g (\vec x,t) & = &H(t)  
\label{eq13}
\end{eqnarray}
but now the observer can think that this field is a gravitational field instead of an inertial effect.

\section{Newtonian cosmology \label{secNC} }
We can implement the Cosmological Principle by simply taking an observer for whom the universe is spatially homogeneous and isotropic. This means that for this observer,  matter has no velocity,  $\vec v =0$, and the gravitational vector potential is $\vec v_g  = 0$. So $\vec g=0$ and $\vec \beta=0$. Under these conditions,  we will see that the generalized Newtonian equations lead to the Friedmann equation for a dust universe.

From equation (\ref{eq12}) we have:
$$
\frac{3}{2}\left(\dot \eta  - \frac{\eta^2 }{2} \right) = 4\pi G\rho 
$$      
Now, from the definition of $\eta$ and defining $H_g(t):= 1/a(t)$ we obtain:
\begin{equation}
\frac{\ddot a}{a} =  - \frac{4\pi G}{3}\rho \label{eq17}
\end{equation}	
And using the mass continuity equation (\ref{eq16g}) with $\vec v=0$:
$$
\frac{\dot \rho }{\rho } = \frac{3}{2}\eta 
$$
and again using the definition of $\eta$:
\begin{equation}
\frac{\dot \rho }{\rho } =  - 3\frac{\dot a}{a}
\label{eq18}
\end{equation}		
We have obtained the Friedmann equations for a dust universe of null curvature. We must point out that this result was obtained using generalized Newtonian mechanics and not General Relativity.

\section{Generalized Newtonian gravitation as the non-relativistic limit of General Relativity\label{sec7}}
If we recognize General Relativity as the ``correct'' theory of gravitation, our generalized Newtonian gravitation should be derivable as its non-relativistic limit. Here we add some new cases to the previous one of \cite{Jaen-Molina}: a spacetime metric $g_{\mu\nu}$ whose geodesics yield the particle equations of motion (\ref{eq9a}) in the limit $c\rightarrow \infty$; and a system of Einstein equations that, in this limit, become  the field equations (\ref{eq11}), (\ref{eq12}).
Let us consider the relativistic extension of the Lagrangian (\ref{eq16b}), 
that is:
\begin{equation}\label{eq34}
L =  - mc^2\sqrt{1-\frac{1}{H_g^2}\frac{(\dot{\vec x} - \vec v_g )^2}{c^2}} \,,
\end{equation}

The action functional is:
$$ S=-mc\int ds =\int L\,dt= -mc\int\sqrt{c^2-\frac{1}{H_g^2}(\dot{\vec x} - \vec v_g )^2}dt$$
This can be obtained through a equivalence principle, i.e., by considering the Minkowski metric in the local system $\mathcal{K}^\prime_{(\vec x,t)}$:
$$
{\bf g^{\prime}}  =  - c^2 dt^2  + d\vec y^2 
$$							
Applying a local transformation of the group of homothetic motions, this metric transforms to:
\begin{equation}\label{eq35}
{\bf g}= -(c^2 -\frac{1}{H_g^2}\vec v_g^2 )dt^2+\frac{1}{H_g^2} d\vec x^2-2\frac{1}{H_g^2}\vec v_g \cdot d\vec x\,  dt
\end{equation}
From now on, we generalize slightly by assuming, because now we are not restricted by the Newtonian equation (\ref{eq12}), that $H_g$ can be a function of space and time.
The metric  (\ref{eq35}) has the following properties:

1) It is form invariant under the group of homothetic motion: $$\left\{ {X^i(t),R_j^i(t),H(t)} \right\},$$  i.e.:
$$
{\bf g} = - c^2 dt^2  + \frac{1}{H^{\prime}_{g}{}^{2}}(d\vec x' - \vec v'_g dt)^2 
$$		
if the fields $\left\{ {H_g(\vec x,t);\;\;\vec v_g(\vec x,t)} \right\}$ transform as:
$$
\vec v'_g = \vec v_g  - \vec v_0; \;\;\;H'_g  = H H_g 
$$						
where $\vec v_0^{} \;$ is given by equation (\ref{eq5})  with $\vec x = \vec X(t) + \vec x'$.\\
From the Einstein equations: 
$$R_{\mu\nu}=\frac{8\pi G}{c^4} t_{\mu\nu}$$
it follows that:

2) If the Newtonian limit, $c\rightarrow\infty$, of the spatial part of the Ricci tensor, $R_{ij}$, which coincides with the Ricci tensor for the three-dimensional metric $ds^2=1/H_g(x,y,z,t)^2(dx^2+dy^2+dz^2)$ must be zero (flat three dimensional space), then, to the lowest order in $1/c$, $H_g(x,y,z,t)\rightarrow H_g(t)$. This equation is equivalent of the field equation for $H_g$, i.e., $\vec\nabla H_g=0$. 

When this last equation is assumed, i.e., $H_g(t)$:

3) The Newtonian limit, $c\rightarrow\infty$, of the particle trajectories that are the geodesics of the spacetime metric (\ref{eq35}) coincides with the generalized Newtonian equations of motion (\ref{eq9a}).

4) The Newtonian limit for the rest of the Ricci tensor components is: 
\begin{eqnarray}
	&&\lim_{c\rightarrow\infty} R_{i0} =\frac{1}{2}\vec\nabla\times \beta \label{eq37a}\\ 
&& \lim_{c\rightarrow\infty} R_{00} =-\vec \nabla \cdot\vec g + \frac{1}{2}\vec \beta ^2 + \frac{3}{2}\left( {\dot \eta  - \frac{{\eta ^2 }}
{2}} \right)\label{eq37}
\end{eqnarray}

If there are no sources of the gravitational field, $R_{\mu\nu}=0$,  these  equations, (\ref{eq37a},\ref{eq37}),  are  the same as those we obtained for Newtonian fields without sources (\ref{eq12}) with $\rho=0$.

5) If the source of the moment energy tensor, $T_{\mu \nu } $, is dust, then:
$$
T_{\mu \nu }  = c^2 \rho u_\mu  u_\nu  
 \Rightarrow t_{\mu \nu }  = c^2 \rho u_\mu  u_\nu   + c^2 \frac{\rho }
{2}g_{\mu \nu } 
$$					
where $u^\mu$ is the unitary velocity of matter:
 $$
u^\mu  = \frac{dx^\mu}{ds} = \frac{\gamma }{c}(1,\vec v);\quad \gamma  \equiv \frac{1}{\sqrt {1 - \frac{1}{H_g^2}\frac{(\vec v - \vec v_g )^2}{c^2} }}
$$				
All the components of $t_{\mu\nu}$ are of order $c^2$ except $t_{00}$ which order is $c^4$,
$$\lim_{c\rightarrow \infty} t_{00}=\frac12 \rho c^4$$
so for the relativistic field equations:
$$
\lim_{c \rightarrow\infty}\left\{  \frac{8\pi G}{c^4} t_{00 }  \right\} = 4 \pi G\rho
$$					
This result, together with (\ref{eq37a},\ref{eq37}), coincides with the Newtonian field equation (\ref{eq12})

6) The velocity field for homothetic motions (\ref{eq5}) together with $H(t)$ can be interpreted as a coordinate change from Minkowski to the metric form (\ref{eq35}) and then the Ricci tensor is zero, i.e., homothetic motions are solutions of the relativistic equation without sources:
\begin{eqnarray}
  && \vec v_0 (\vec x,t) = \dot{\vec X}(t) + \vec \Omega (t) \times (\vec x - \vec X(t)) + \frac{{\dot H}}
{H}(\vec x - \vec X(t))  \\
  && H_0 (\vec x,t) = H(t) 
\end{eqnarray}

\section{FLRW spacetime\label{sec8}}
Now we look at the FLRW metric \cite{Lemaitre}, \cite{Friedmann}, \cite{Robertson}, \cite{Walker} to study the corresponding relativistic metric form (\ref{eq35}), the notions of homothetic motion  and its limit to Newtonian cosmology. 
Using isotropic coordinates, the FLRW metric is:
$$
{\bf g}  =  - c^2 dt^2  + \frac{a(t)^2}{(1 + \frac{k}{4}\vec x^2)^2}d\vec x^2 
$$	
which can be compared with metrics of our type:
\begin{equation}
{\bf g}  =  - c^2 dt^2  + \frac{1}{H_g^2}(d\vec x  - \vec v_g dt)^2
\label{eq20}
\end{equation}
\begin{equation}
H_g = \frac{1 + \frac{k}{4}\vec x^2}{a(t)}; \quad \vec v_g  = 0
\label{eq21}
\end{equation}
We can choose a homothetic frame: 
$$
\vec x = \frac{1}{a}\vec y \Rightarrow d\vec x = \frac{1}{a}(d\vec y - \frac{\dot a}{a}\vec ydt)
$$		
such that for this observer we have the same metric (\ref{eq20}) but now:
\begin{equation}
H_g = 1 + \frac{k}{4}\left(\frac{\vec y}{a}\right)^2; \quad \vec v_g  = \frac{\dot a}{a}\vec y
\label{eq22}
\end{equation}
As the function $H_g$ depends on $\vec y$, then this metric does not have a good Newtonian limit unless $k=0$:  flat 3-space.
The best frame for a good non-relativistic limit is (\ref{eq22}), although for this observer the velocity of the matter is not zero, i.e., it is not a comoving observer. In this frame, the Newtonian limit, for $k=0$ and when we have dust matter, is the Newtonian cosmology of Section \ref{secNC} above.

Spaces with $k=1,\; k=-1$ can be incorporated if we assume that $a(t)=c/u(t)$, where $u(t)$ has the dimensions of velocity, and then:
\begin{equation}
H_g = 1 + \frac{k}{4}\left(\frac{u(t)\vec y}{c}\right)^2; \quad \vec v_g  = -\frac{\dot u}{u}\vec y
\label{eq23}
\end{equation}
so the curvature of space appears of the order $1/c^2$, and the Newtonian limit is always flat.
\section{Conclusions}
In this paper we follow an  procedure  that is an extended version of \cite{Jaen-Molina}. We have characterized the homothetic group of motion by enlarging the rigid motion group. After analyzing  the inertial force fields, inertial potentials and the equations fulfilled by these fields, we used these kinds of potential to define gravitational fields.  In this way we arrive at one of the main goals of the paper: to set up the theory of Newtonian gravitation in a homothetic invariant formulation.

Using this theory we have been able to apply the Cosmological Principle, at a Newtonian level, and find the Friedman equation for a dust universe without using General Relativity at all: the second main goal of the work.

By writing the equation of motion for the particles by means of a Lagrangian formulation we found its relativistic version and we provide a spacetime metric that has the gravity potentials as its unique ingredient. 

This spacetime metric has many interesting  properties. Among them are the fact that it is covariant under homothetic transformations and that the corresponding Einstein field equations for dust has a non-relativistic limit which leads to the  homothetic covariant Newtonian gravitation. We want to emphasize that in this limiting process there are no considerations other than the limiting procedure $c\rightarrow\infty$, i.e., there is no need for a weak field approximation. This is the third main goal of the paper.

The family spacetime metrics defined here includes the FLRW metric. This allows us, by performing homothetic transformation and taking into account its non-relativistic limit, to reproduce the Newtonian cosmology we introduced in \ref{secNC}.

\section*{Acknowledgments}
We thank Josep Llosa for carefully reading the paper and for providing useful criticism that has led to improvements.

\end{document}